\documentclass{aa}
\usepackage{graphicx}
\usepackage{txfonts}
\usepackage{natbib}
\usepackage{amssymb}
\usepackage{amsmath}
\usepackage{longtable}
 \usepackage[normalem]{ulem}
\bibpunct{(}{)}{;}{a}{}{,}
\def \kms{km~s$^{-1}$}
\def \fwhmR{${\rm FWHM_{res}}$}
\def \Tgas{${\rm T_{gas}}$}
\def \nodata{. . .}
\def \bturb{$b_{\rm turb}$}
\renewcommand{\ion}[1]{~\textsc{#1}}

\begin{document}

   \title{Performance of ESPRESSO's high resolution 4x2 binning for characterizing intervening absorbers towards faint quasars}

   \titlerunning{ESPRESSO HR42 performance for observing quasars}
   \authorrunning{T.A.M.~Berg et al.}

   \author{Trystyn A. M. Berg
          \inst{1}\fnmsep\inst{2}\fnmsep\thanks{tberg@eso.org}
          \and
          Guido Cupani\inst{3}
          \and
          Pedro Figueira\inst{1}\fnmsep\inst{4}
          \and
          Andrea Mehner\inst{1}
          }
   \institute{
            European Southern Observatory, Alonso de Cordova, 3107, Casilla, 19001, Santiago, Chile
         \and
             Departamento de Astronom\'ia, Universidad de Chile, Casilla, 36-D, Santiago, Chile
        \and
            INAF-Osservatorio Astronomico di Trieste, Via Tiepolo 11, I-34143, Trieste, Italy
        \and
            Instituto de Astrof\'{i}sica e Ci\^{e}ncias do Espa\c{o}, Universidade do Porto, CAUP, Rua das Estrelas, 4150-762, Porto, Portugal}



  \abstract
   {
   As of October 2021 (Period 108), the European Southern Observatory (ESO) offers a new mode of the ESPRESSO spectrograph designed to use the High Resolution grating  with $4\times2$ binning (spatial by spectral; HR42 mode) with the specific objective of observing faint targets with a single Unit Telescope at Paranal. We validated the new HR42 mode  using four hours of on-target observations of the quasar J0003-2603, known to host an intervening metal-poor absorber along the line of sight. The capabilities of the ESPRESSO HR42 mode (resolving power $R\approx137~000$) were evaluated by comparing to a UVES spectrum of the same target with a similar integration time but lower resolving power ($R\approx48~000$). For both  data sets we tested the ability to decompose the velocity profile of the intervening absorber using Voigt profile fitting and extracted the total column densities of C\ion{iv}, N\ion{i}, Si\ion{ii}, Al\ion{ii}, Fe\ion{ii}, and Ni\ion{ii}.  With $\approx3\times$ the resolving power and $\approx2\times$ lower S/N for a nearly equivalent exposure time, the ESPRESSO data is able to just as accurately characterize the individual components of the absorption lines as the comparison UVES data, but has the added bonus of identifying narrower components not detected by UVES. For UVES to provide similar spectral resolution ($R>100~000$; 0.3$''$ slit) and the broad wavelength coverage of ESPRESSO, the Exposure Time Calculator (ETC) supplied by ESO estimates 8~hrs of exposure time spread over two settings; requiring double the time investment than that of ESPRESSO's HR42 mode whilst not properly sampling the UVES spectral resolution element. Thus ESPRESSO's HR42 mode offers nearly triple the resolving power of UVES (0.8$''$ slit to match typical ambient conditions at Paranal) and provides more accurate characterization of quasar absorption features for an equivalent exposure time.

   }


   \maketitle

\section{Introduction}

Gaseous systems seen in absorption along quasar sightlines are excellent probes of astrophysics throughout cosmic time, from measuring the physical and chemical evolution of gaseous reservoirs of galaxies \citep{Berg15,Neeleman15,Noterdaeme21} to imposing constraints on dark matter properties \citep{Irsic17} and fundamental constant evolution in our Universe \citep[e.g.,][]{Murphy17,Schmidt21}. Accurately quantifying these probes requires a detailed kinematic decomposition of all absorption lines along a quasar sightline. For distinct QSO absorption line systems  (QALs),  the metal absorption profiles are often quite complex, consisting of several sub-components (with a typical width between $\approx5-100$~\kms{}) spread out over a few hundred \kms{}. In addition, the neutral hydrogen absorption of individual absorbers are sometimes significantly blended within the Ly$\alpha$ forest absorption from the intervening intergalactic medium. While low resolution spectra (with resolving power $R<1000$) can easily identify the strongest Ly$\alpha$ QALs such as damped Lyman $\alpha$ systems \citep[DLAs; H\ion{i} column densities of $\geq 2\times10^{20}~{\rm atom~cm^{-2}}$;][]{Wolfe05}, spectrographs like UVES \citep[][]{Dekker00} are frequently used to  identify the weaker H\ion{i} absorbers, which are required to probe the intergalactic and circumgalactic media. Typical studies of faint quasars with UVES use a $1''$ slit ($R\approx48~000$) to maximize $R$ while minimizing light loss based on the median seeing of Paranal ($\approx0.8''$).

Although UVES has proved to be an effective instrument for the study of QALs over the past 20 years, the QAL community are now exploring higher spectral resolving power ($R\gtrsim100~000$; i.e., the smallest cloud widths of $\gtrsim3$~\kms{}) to pinpoint the detailed structure of the absorption line profiles, and search for rarely detected species, such as:  (i) deuterium and lithium for constraining Big Bang nucleosynthesis \citep{Cooke18}, (ii) molecular hydrogen to constrain the gas and dust properties of high redshift galactic gas reservoir \citep{Tchernyshyov15, deCia18, Krogager18}, and (iii) exotic nucleosynthetic species to place constraints on chemical evolution processes \citep{Ellison02,Berg13}. To do these experiments with UVES would require the smallest slit widths (0.3$''$; thus $R\approx100~000$), leading to large amounts of light loss at the slit interface unless observations are completed in the most pristine atmospheric seeing conditions. In essence, it is not efficient for UVES to continue to explore the emerging questions of the QAL community.

The ESPRESSO spectrograph \citep{Pepe10,Pepe21} provides an excellent alternative to UVES for observing QALs at high spectral resolution.  In its medium-resolution mode, the 4 Unit Telescope (UT) MULTIMR42 mode ($R=75~000$ with 4x2 binning), ESPRESSO has already proven to be effective at using QALs to constrain our understanding of the Universe \citep{Cooke20,Welsh20}. However, these observations require all 4 UTs simultaneously, and the collection of light from MULTIMR42 mode means a reduction of $R$ from $\approx 137\,000$ to $75\,000$; a resolution UVES can obtain without excessive slit losses in  average Paranal atmospheric conditions. As of Period P108, ESPRESSO supports a single UT mode with $4\times2$ binning (HR42 mode). This mode is designed to reduce read-out noise while maintaining a high spectral resolving power ($R\approx137~000$) and is ideal for observing faint targets such as QSOs.

In this paper, we present the results from a pilot program to test the ability of ESPRESSO's HR42 mode to analyze the chemical contents of a DLA seen along a quasar sightline, and compare the performance of ESPRESSO HR42 mode to the typical observing set-up used for UVES.

\section{Target selection, observations and data reduction}

The target, QSO~J0003$-$2603 (right ascension 00~hr $03'$ $22.953''$, declination $-26^\circ$ $03'$ $18.307''$, and redshift $z_{\rm em}=4.01$), was chosen as it is: ($i$) sufficiently bright for the ESPRESSO autoguiding system (AB magnitude $R = 18.5$~mag),  ($ii$) known to host an intervening metal-poor ([M/H]~$\approx-2$) DLA absorber along the line of sight (redshift $z_{\rm abs}=3.39$, and ($iii$) has archival UVES observations allowing a direct comparison between instruments \citep[][]{Molaro00,Levshakov00}.

The ESPRESSO observations were carried out on October 2$^{\rm nd}$ 2021 on UT2 at Paranal (Program ID 60.A-9801(W)). Four sequential 60-minute on-target exposures were taken. The observations were made in average seeing conditions ($\approx0.77''$; image quality at 5000~\AA{} of $\approx0.85''$) across the four hours of observations at an average airmass of 1.05. The data were reduced with the default settings of the ESPRESSO ESO-reflex pipeline \citep[version 2.3.3;][]{Pepe21,Freudling13} and post-processed with the ESPRESSO data analysis software \citep[hereafter DAS; version 1.3.3;][]{Cupani19,Pepe21}. While we use the sky-subtracted data in this paper to avoid skyline contribution, we point out that the S/N is marginally better in the spectrum without sky subtraction. The S/N (in units of pixel$^{-1}$; $\approx1$~\kms{} per pixel) is $\approx20$ within the Ly$\alpha$ forest, and $\approx30$ beyond the Ly$\alpha$ emission line of the quasar.

The quoted resolving power of the HR42 mode, based on the ThAr arc calibrations taken by ESO, is $R\approx131~000$. To compare what was achieved by our data, we quantified $R$ using the telluric lines from two different methods. We first assumed that the measured full width half maximum (FWHM) of a weak, unresolved telluric line is equal to the FWHM of the resolution element (\fwhmR{}); the measured \fwhmR{} of the telluric line at $\approx7685$~\AA{} within the median-combined spectrum is $2.269\pm0.292$~\kms{} (corresponding to $R\approx132~100$). However, this method assumes that telluric lines are unresolved by ESPRESSO, which is not necessarily true given the high spectral resolution of the HR42 mode. To confirm the specific telluric line was unresolved and to obtain a second, independent measurement of the resolving power, we fitted the telluric lines using using \textsc{molecfit} \citep{Smette15}, which provides both an estimate of \fwhmR{} as well as the intrinsic FWHM of the telluric line based on a model atmosphere at the time of observations. Taking the mean value from each individual exposure, the \textsc{molecfit}-estimated intrinsic FWHM of the telluric line at $\approx7685$~\AA{} is $1.480\pm0.050$~\kms{} and \fwhmR{}~$=2.193\pm0.050$~\kms{} ($R\approx136~700$). It is clear from the results of \textsc{molecfit} that the telluric line at $\approx7685$~\AA{} is marginally unresolved within our dataset. However we caution that depending on the airmass and atmospheric conditions at time of observations, the telluric lines may be resolvable by ESPRESSO's HR mode and will not provide an accurate estimate of the resolving power of the instrument. We adopt $R\approx136~700$ from \textsc{molecfit} as the best estimate of the resolving power of the HR42 mode.

The comparison archival UVES data used 4.5 hours of on-target exposures, and was combined and continuum normalized by the UVES Spectral Quasar Absorption Database team \citep[SQUAD;][]{Murphy19}. Observations were executed during the commissioning of UVES (Program ID 60.A-9022(A); slit width 0.9$''$ with the DIC2 437+860 instrument setup), and were observed at similar airmass and at various seeing measurements (from 0.34$''$ to 2.4$''$; mean of $1.0''$ at 5000~\AA{}) across two nights. Using unresolved telluric lines within the data, the obtained FWHM of the spectral resolution element is \fwhmR{}~$=5.981\pm0.126$~\kms{} ($R\approx50~100$; archival $R$ measurements from ThAr arc lines obtained $R\approx47~900$). The UVES pixel size is $\approx2.5$~\kms{}. The S/N  per pixel is $\approx2\times$ (alternatively, $\approx 1.33\times$~per \AA{}) larger within the UVES spectrum relative to that of ESPRESSO (Figure \ref{fig:snr}). Once accounting for the difference in pixel size between the two instruments, the S/N of UVES becomes significantly better than ESPRESSO at $\lesssim5000$~\AA{}.

\section{Results}
\subsection{Comparison with the ESO Exposure Time Calculator}

\begin{figure*}
    \centering
    \includegraphics[width=\textwidth]{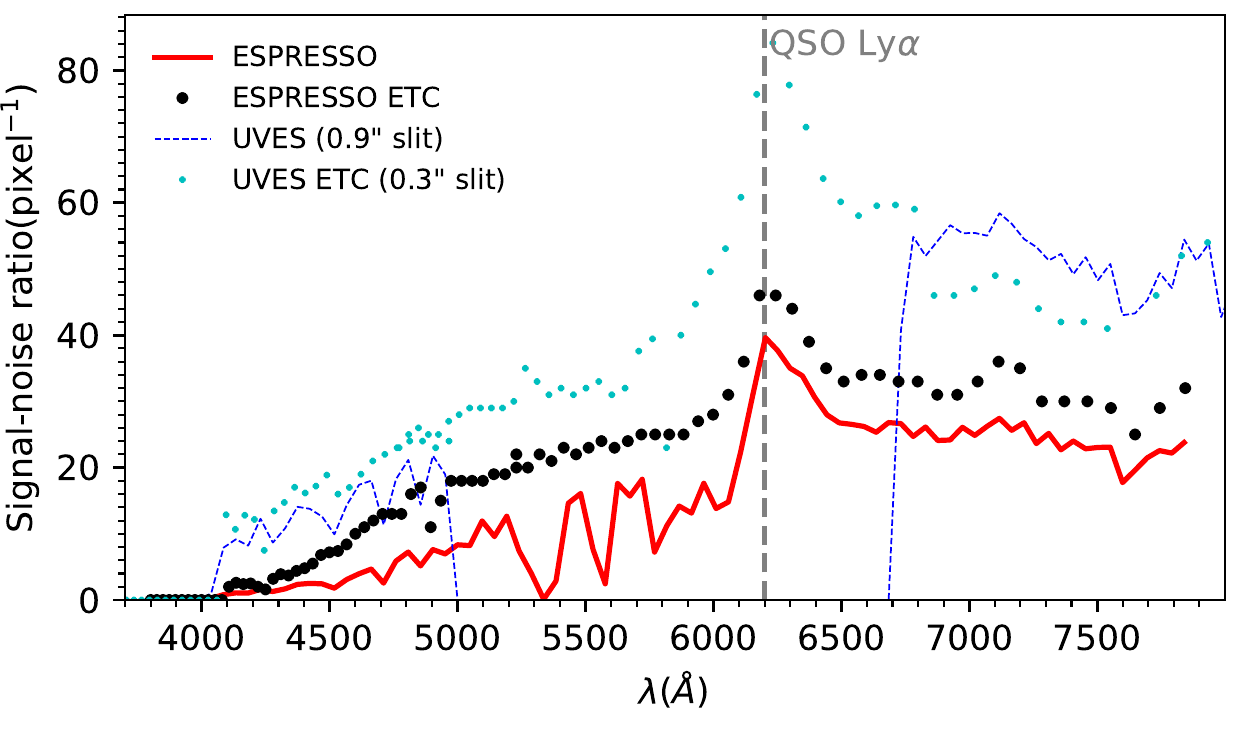}
    \caption{The S/N (per pixel; $\approx 1$~\kms{} per pixel) of the observed ESPRESSO spectrum (solid red line) and ETC-estimated values based on a QSO emission template (black circles). The observed spectrum S/N is estimated as the measured flux divided by the standard deviation from the error spectrum. Note that the red line is smoothed by taking the median S/N within a wavelength bin of the same size as the average space between ETC points. For reference, the vertical grey line denotes the Ly$\alpha$ emission line of the observed QSO. The S/N from the ETC consistently $\lesssim20$~percent larger than measured in the observations. For reference, the smoothed S/N from the comparison UVES data ($\approx2.5$~\kms{} per pixel) is shown as the blue dotted line. Due to the gaps in the UVES data wavelength coverage, the observed UVES data does not cover the Ly$\alpha$ emission of the QSO. Additionally, the small cyan dots represent the combined ETC estimate for UVES data taken with two UVES settings, with each setting using a 0.3$''$ slit (near equivalent $R$) with the same exposure time \emph{per setting} (i.e., doubled observing time) and observing conditions as the ESPRESSO data.}
    \label{fig:snr}
\end{figure*}

Figure \ref{fig:snr} shows the observed S/N (red line) across the entire spectrum compared to what was estimated using version P109 of the ESPRESSO exposure time calculator (ETC; black points). The ETC S/N estimates have been computed assuming a $0.84''$ image quality, the average airmass through the observations (1.05), and fractional lunar illumination FLI=0.0 as the moon was not visible during observations. The S/N estimated by the ETC is consistently $\approx1.3\times$ higher across the entire wavelength range. However, we point out that the ETC solely includes emission from the QSO continuum and not absorption lines observed towards QSO J0003$-$2603 (particularly those of the Ly$\alpha$ forest; $<6202$~\AA{}), thus overestimating the estimated S/N. This is particularly demonstrated by the presence of the strong Ly$\alpha$ absorption from the DLA at $z_{\rm abs}=3.39$ ($\approx5300$~\AA{} in red line of Figure~\ref{fig:snr}). Using only pixels of the observed ESPRESSO spectrum with predominantly quasar continuum flux near the centers of ESPRESSO's spectral orders, the observed S/N is only $\approx1.15\times$ smaller than the ETC predicts.

To obtain a near-equivalent spectrum with UVES (i.e., $R\approx110~000$) and wavelength coverage from 3800~\AA{} to 7880~\AA{}), one would need to observe with two UVES settings (e.g., DIC1 390+580 and DIC2 437+760) and with slit widths of 0.3$''$. The cyan points in Figure \ref{fig:snr} show the ETC-predicted S/N for the UVES spectrum using these two settings and 0.3$''$ slit width, with an exposure time of four hours \emph{per setting} under the same observing conditions as the ESPRESSO  observations. The S/N of the cyan points is $\approx 2\times$ higher per pixel than what is observed with ESPRESSO for two reasons. First, as there is overlap in wavelength between the two UVES settings, there is an increase in the S/N in the wavelength ranges $\approx$3760-4480~\AA{},  $\approx$4770-4960~\AA{}, and $\approx$5700-6780~\AA{}. Secondly, the UVES unbinned pixel sampling of the resolution element of this setup is 1.39 pixels per element in the blue and 1.94 pixels in the red. Not only does this pixel size lead to under-sampling in the spectral direction, but also it results in an increase in S/N per pixel of 1.18--1.39$\times$ compared to ESPRESSO. Thus the expected ratio of the S/N between UVES and ESPRESSO -- after accounting for the doubled exposure time in overlapping regions of the two UVES settings, and the resolution element size -- is between 1.18 and 1.97, and explains the majority of the discrepancy between the red curve and cyan points of Figure \ref{fig:snr}. While ESPRESSO is expected to be as efficient as UVES with a 0.3" slit to obtain a $R>100~000$ spectrum, ESPRESSO is able to obtain full wavelength coverage (3800\AA{} to 7880\AA{}) in a single setup whilst Nyquist sampling the spectral resolution element, making the HR42 mode more efficient in terms of observing time required.

\subsection{Characterization of the intervening absorber with ESPRESSO}

\begin{figure}
    \centering
    \includegraphics[width=0.5\textwidth]{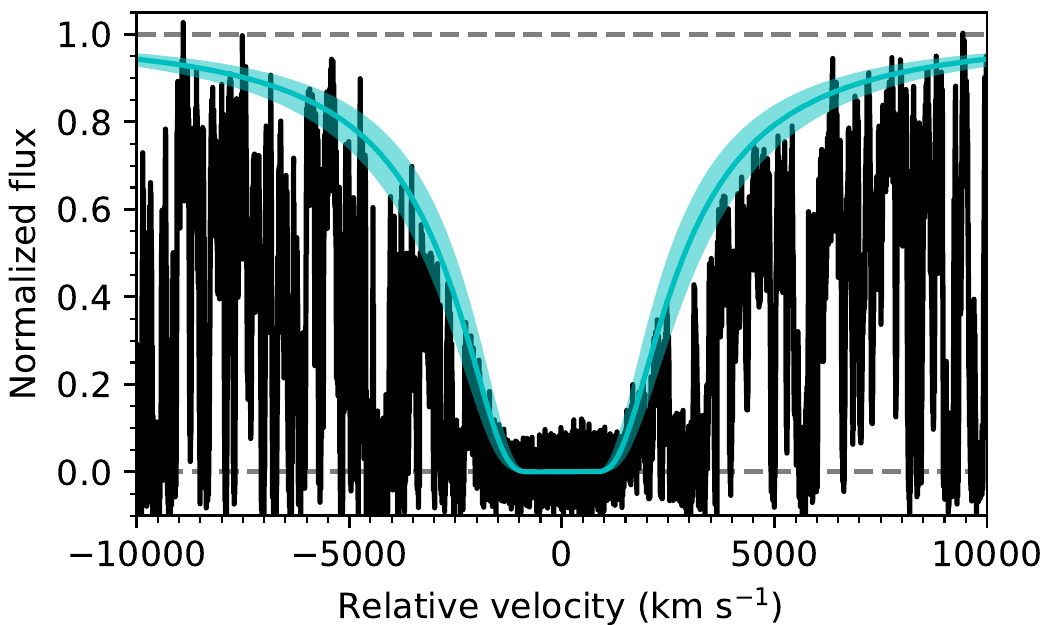}
    \caption{The best-fit Voigt profile of Ly$\alpha$ (cyan line) to the continuum-normalized ESPRESSO data (black line) for the absorption seen at $z_{\rm abs}=3.3901$. The best fitting H\ion{i} column density is logN(H\ion{i})~$=21.35\pm 0.12$. The shaded cyan region denotes the range of profiles from the error on logN(H\ion{i}). }
    \label{fig:HI}
\end{figure}

By simultaneously fitting the continuum and a Voigt profile to both the Ly$\alpha$ and Ly$\beta$ lines, the measured H\ion{i} column density of the  intervening absorber towards QSO~J0003$-$2603 is logN(H\ion{i})~$=21.35\pm 0.12$ at $z_{\rm abs}=3.3901$. The measured H\ion{i} column density is consistent with previous measurements \citep{Lu96}. The Voigt profile fit to the Ly$\alpha$ absorption line is shown in Figure \ref{fig:HI}.

Additionally, Voigt profile fitting was done for absorption lines of key metal species (C\ion{iv}, N\ion{i}, Si\ion{ii}, Al\ion{ii}, Fe\ion{ii}, Ni\ion{ii}). We note that in the analysis of the UVES data done by \cite{Molaro00}, there was sufficient S/N in the Ly$\alpha$ forest ($\ge10$~pixel$^{-1}$) to detect the O\ion{i} lines $\lambda$ 925~\AA{} and 950~\AA{} ($\approx 4115$~\AA{} in the observed frame). However, the efficiency of ESPRESSO at blue wavelengths is too poor to be able to accurately model the continuum within the Ly$\alpha$ forest, although it is possible to see evidence of saturation of the O\ion{i} $\lambda$ 929~\AA{} and 950~\AA{} lines. Unfortunately, the wavelength coverage of ESPRESSO does not include coverage of  Cr\ion{ii} and Zn\ion{ii} at the redshift of the DLA.

\begin{figure*}
\centering
\includegraphics[width=\textwidth]{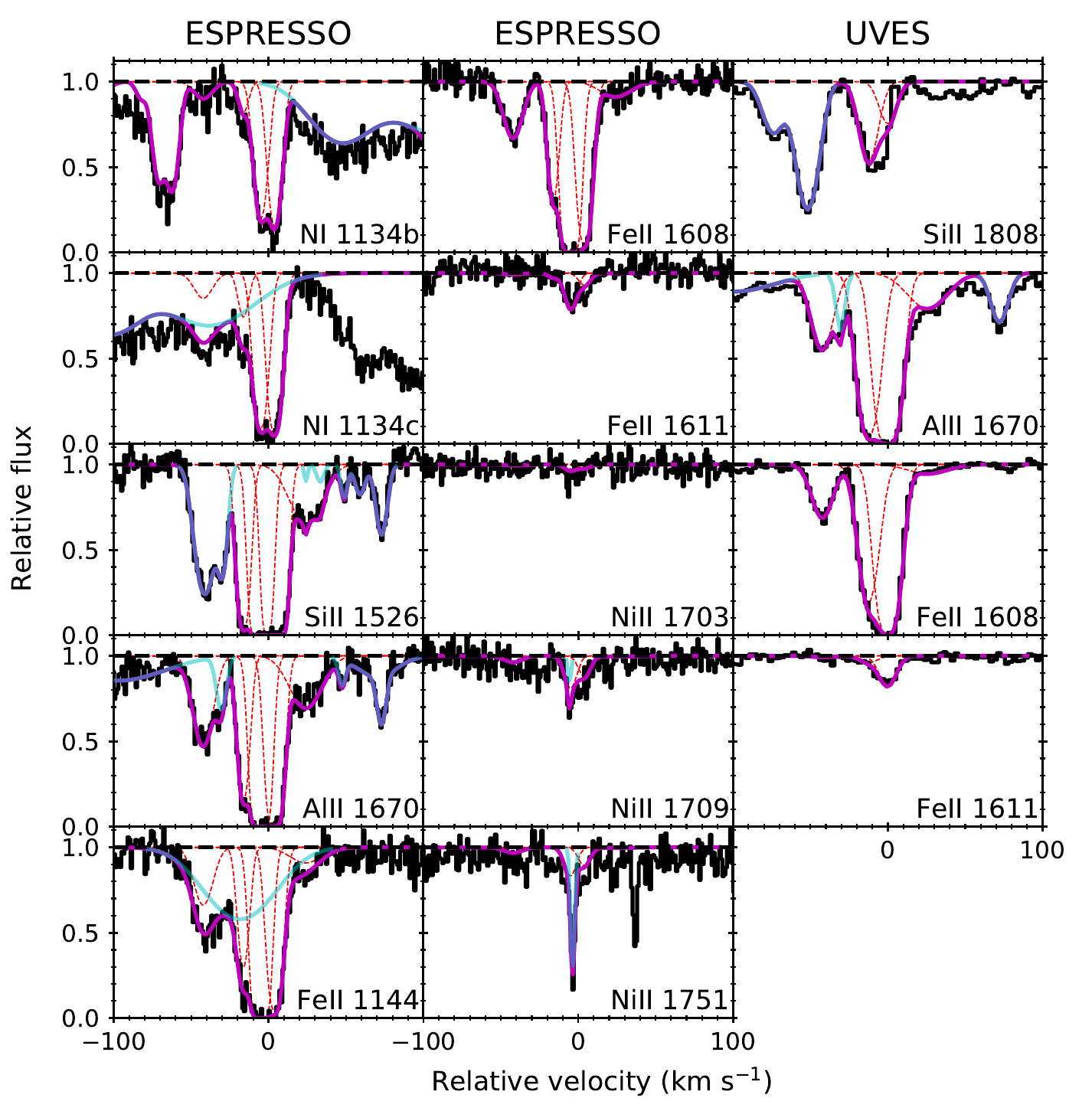}
\caption{Velocity profiles of key low-ionization absorption lines (labelled in the bottom right corner of each panel) detected within the continuum-normalized data (black lines) of QSO~J0003$-$2603. The panels of the left two columns of the figure denote data from ESPRESSO, while the rightmost column shows the absorption lines of the UVES data. The solid magenta curve denotes the complete Voigt profile fitted to the data, while the red dashed lines denote the components that constitute the fit. Overplotted in cyan are the blended components. Although the profile for N\ion{i} $\lambda$ 1134.1~\AA{} was not fit due uncertain continuum blueward of the line, the expected profile shape based on the fits of the other two triplet lines ($\lambda$ 1134.4~\AA{} [1134b] and $\lambda$ 1135.0~\AA{} [1134c]) is shown in the N\ion{i} 1134b panel at $\approx-70$~\kms{}.}
\label{fig:lowions}
\end{figure*}

The fitting was done using the code \textsc{ALIS}\footnote{\url{https://github.com/rcooke-ast/ALIS}}, which uses a $\chi^{2}$ minimization to simultaneously fit the continuum and profile components from a list of starting redshifts ($z$), the turbulent component  of the Doppler broadening parameters (\bturb{}), gas temperature (\Tgas{}), and column densities (N; in units atoms cm${-2}$) for each ionic species. A Gaussian FWHM of 2.193 \kms{} is used as the ESPRESSO line spread function in the fitting procedure. As the HR42 mode provides sufficient resolution to possibly constrain gas temperature \citep{Narayanan06}, we attempted including \Tgas{} as a free parameter per absorption component in the fitting process. Despite having a range of ions of varying atomic mass, many lines in this system  are blended or saturated preventing a realistic constraint on \Tgas{}; including \Tgas{} as a free parameter in the model suggests a purely turbulent medium. In order to obtain an accurate measure of the column densities for all components of each ion, we elect to assume \Tgas{}~$=10~000$~K, a typical temperature of metal-poor DLAs \citep{Cooke15} rather than separately fitting a single Doppler parameter to each ions' component. While assuming a single \Tgas{} may not provide accurate measurements of \bturb{} for each component, the Voigt profile fitting procedure will be more precise as the fit is constrained by more data (i.e.~multiple ions' probing the same \bturb{}) and effects from atomic mass are properly accounted for. As a result, this will reduce the overall errors in both the Doppler parameter and logN. Furthermore, by deriving \bturb{} for each component, the fitting also breaks the degeneracy between the Doppler parameter and column density for saturated lines, enabling a measurement of the column density for Si\ion{ii} and Al\ion{ii}. We note that repeating the fitting procedure with a \Tgas{} within the range of DLA values (i.e.~between 1000K -- 100~000K) has little effect on the derived column densities, which remain consistent with the assumed \Tgas{}$=10~000$~K column densities. By assuming a fixed \Tgas{}, the errors derived from fitting \bturb{} encapsulate the errors on the total Doppler parameter. The high ions (i.e., C\ion{iv}) and low ions (N\ion{i}, Si\ion{ii}, Al\ion{ii}, Fe\ion{ii}, and Ni\ion{ii}) were fitted separately, and assuming that the $z$ of each component of the low ions' velocity profile are the same across all species. Given the level of blending and saturation for many of these lines which introduces model degeneracies while fitting all lines simultaneously, the velocity profiles were fitted in several iterations to constrain individual components free of these degeneracies. First, the strongest components at $\approx0$~\kms{} were fitted using the unsaturated Fe\ion{ii} $\lambda$ 1611~\AA{} and the two redmost lines of the N\ion{i} $\lambda$ 1134~\AA{} triplet  ($\lambda$ 1134.4 and 1135.0~\AA{}) to fix \bturb{}, $z$ and N(Fe\ion{ii}) for these two components that cannot be easily determined from the saturated lines. The redshifts of the rest of the profile (five components total) were then fixed using the unblended Fe\ion{ii} $\lambda$ 1608~\AA{}, before running the fit on the remaining lines to determine the remaining parameters. These ALIS models were subsequently cross-checked with the ESPRESSO DAS; the results of the two methods are in agreement within the uncertainties.  The best-fit absorption profile parameters are presented in the top of Table \ref{tab:loComp}, while the velocity profiles and their corresponding fits are shown in the two leftmost columns of Figure \ref{fig:lowions}. For reference, the  velocity of each component relative to $z_{\rm abs}$ ($\Delta v$) is tabulated in Table \ref{tab:loComp}.

 Similarly, we applied the same profile fitting method to the low ions in the UVES data (assuming an instrumental resolution element of 5.981 \kms{}; right column of Figure \ref{fig:lowions}; bottom of Table \ref{tab:loComp}), and detected only four components. Components 2, 3 and 4 of the ESPRESSO fit (Table \ref{tab:loComp}) are separated by $\approx20$~\kms{}, and all three components have a \bturb{} below the resolving power of the 0.9$''$ UVES set-up (FWHM~$\approx6.4$~\kms{}). As component 2 is weaker than the other two components, the asymmetry in the line profile can be detected within the UVES data, while components 3 and 4 cannot be distinguished. Thus, as a result of the $\approx3\times$ larger spectral resolving power of the HR42 mode, more components with small \bturb{} can be detected with ESPRESSO despite the modest S/N.

 Given the strength of the Si\ion{ii} $\lambda$ 1526 line in the ESPRESSO data, we point out that there appears to be an artefact in the UVES data near the Si\ion{ii} $\lambda$ 1808~$\AA$ line. The profile shape appears shifted blueward in velocity space ($\approx -5$~\kms{}) with reference to the other lines (right column of Figure \ref{fig:lowions}), leading to a poor fit using the fixed redshift and \bturb{} from the Al\ion{ii} and Fe\ion{ii} lines. The resulting logN(Si\ion{ii})  is severely inconsistent with the ESPRESSO value. Fitting a separate, independent two-component profile to the Si\ion{ii} 1808 line results in a total column density estimate of logN(Si\ion{ii})~$=15.65$, which is inconsistent with the sum of the components 2, 3 and 4 of the equivalent absorption feature in the ESPRESSO data (logN(Si\ion{ii})~$=15.07\pm0.16$).

 \begin{table*}
 \small
\begin{center}
\caption{Component Voigt profile fits for low ions}
\label{tab:loComp}
\begin{tabular}{lcccccccc}
\hline
Comp.& $z$& $\Delta v$ (km s$^{-1}$)& \bturb{} (km s$^{-1}$)& logN(N\ion{i})& logN(Si\ion{ii})& logN(Al\ion{ii})& logN(Fe\ion{ii})& logN(Ni\ion{ii})\\
\hline
\multicolumn{9}{c}{\textbf{ESPRESSO}}\\
1& $3.389512$& $-42$& $8.1\pm0.27$& $13.25\pm0.175$& \nodata{}& $12.14\pm0.017$& $13.38\pm0.058$& $12.59\pm0.271$\\
2& $3.389892$& $-16$& $4.3\pm0.09$& $13.30\pm0.082$& $13.71\pm0.068$& $12.30\pm0.017$& $13.59\pm0.015$& \nodata{}\\
3& $3.390056$& $-5$& $5.1\pm0.12$& $14.26\pm0.021$& $14.56\pm0.198$& $13.17\pm0.068$& $14.57\pm0.026$& $13.14\pm0.045$\\
4& $3.390188$& $4$& $5.1\pm0.09$& $14.32\pm0.019$& $14.73\pm0.053$& $12.86\pm0.030$& $14.07\pm0.015$& $12.92\pm0.052$\\
5& $3.390481$& $24$& $13.7\pm0.68$& \nodata{}& $13.29\pm0.028$& $12.06\pm0.024$& $12.97\pm0.062$& $11.96\pm0.911$\\
\hline
\multicolumn{9}{c}{\textbf{UVES}}\\
1& $3.389511$& $-42$& $8.4\pm0.39$& \nodata{}& \nodata{}& $12.09\pm0.054$& $13.40\pm0.019$& \nodata{}\\
2& $3.389949$& $-12$& $7.0\pm0.11$& \nodata{}& $14.96\pm0.009$& $12.79\pm0.012$& $13.99\pm0.008$& \nodata{}\\
3& $3.390130$& $0$& $6.5\pm0.09$& \nodata{}& $14.59\pm0.018$& $13.29\pm0.029$& $14.63\pm0.018$& \nodata{}\\
4& $3.390505$& $26$& $15.4\pm1.17$& \nodata{}& \nodata{}& $11.92\pm0.035$& $12.82\pm0.086$& \nodata{}\\

\end{tabular}
\end{center}
\end{table*}

\begin{figure*}
\centering
\includegraphics[width=\textwidth]{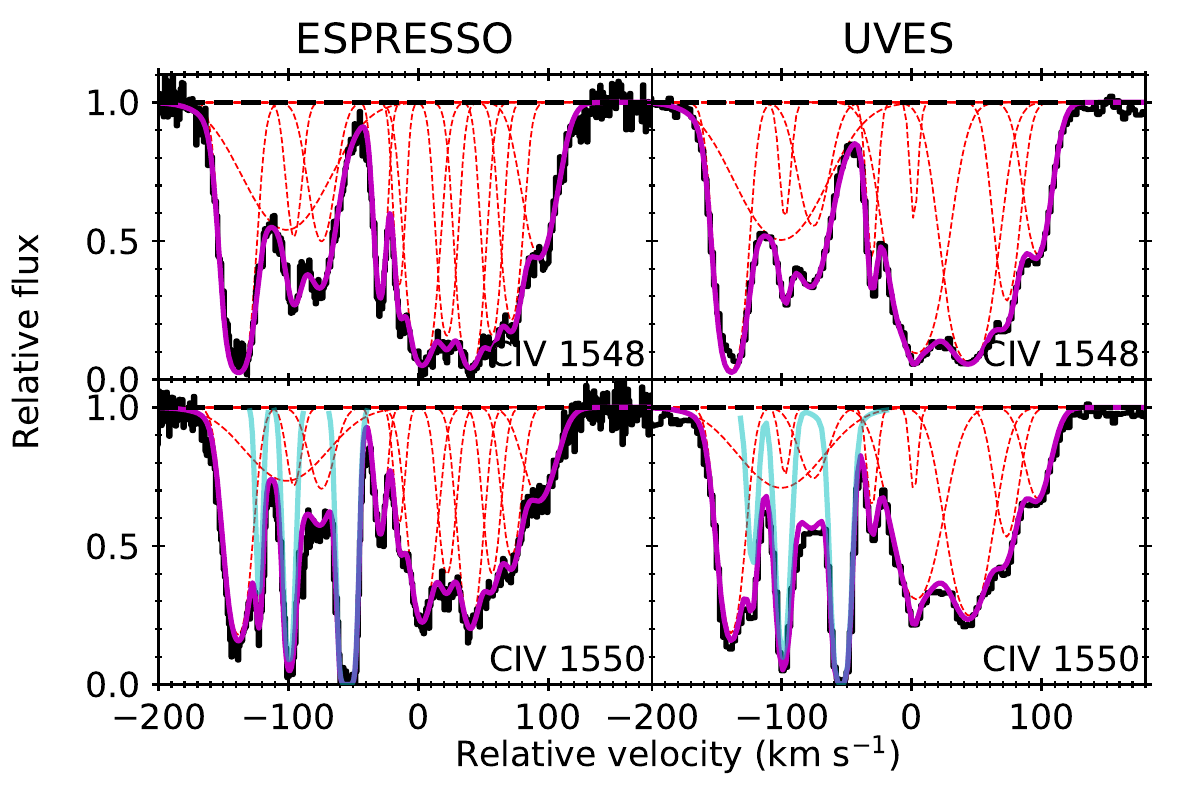}
\caption{ Velocity profiles of the C\ion{iv} doublet absorption seen in the ESPRESSO (left column) and UVES (right column) data of QSO~J0003$-$2603. Notation is the same as in Figure \ref{fig:lowions}.}
\label{fig:CIV}
\end{figure*}

The same fitting method was also applied to the C\ion{iv} doublet velocity profiles  (Figure \ref{fig:CIV}). The complexity of the C\ion{iv} absorption profile combined with either the modest S/N of the ESPRESSO or the resolution of the UVES data leads to a degenerate parameter space, requiring anywhere between 10 and 16 components to provide a satisfactory fit. Many of the additional components beyond the initial 10 are much weaker, with either the column density being several orders of magnitude smaller than the total value and thus consistent with the noise, or the components are shallow and/or broad in shape and could be accounted for by modifying the continuum fit. In many cases, these additional components contribute minimally to the total column density. For the UVES data, the minimum 10 components were sufficient to reproduce the absorption profile (right column of Figure \ref{fig:CIV}). As with the low ion profile, the ESPRESSO data is able to resolve an additional two components (at $\Delta v = 22$ and 57~\kms{}; left column of Figure \ref{fig:CIV}).  The properties of each component for both datasets are tabulated in Table \ref{tab:hiComp}. The recovered total column density of C\ion{iv} from UVES and ESPRESSO are equivalent.

\begin{table*}
\begin{center}
\caption{Component Voigt profile fits for high ions}
\label{tab:hiComp}
\begin{tabular}{lcccc}
\hline
Comp.& $z$& $\Delta v$ (km s$^{-1}$)& $b$ (km s$^{-1}$)& logN(C\ion{iv})\\
\hline
\multicolumn{5}{c}{\textbf{ESPRESSO}}\\
1& $3.388100$& $-139$& $11.5\pm0.17$& $13.96\pm0.017$\\
2& $3.388645$& $-101$& $41.4\pm1.63$& $13.76\pm0.023$\\
3& $3.388729$& $-96$& $7.5\pm0.55$& $13.06\pm0.039$\\
4& $3.389040$& $-74$& $12.8\pm0.78$& $13.31\pm0.036$\\
5& $3.389703$& $-29$& $5.9\pm0.18$& $13.22\pm0.012$\\
6& $3.389918$& $-15$& $5.4\pm0.44$& $13.13\pm0.067$\\
7& $3.390172$& $3$& $11.9\pm1.04$& $13.91\pm0.035$\\
8& $3.390453$& $22$& $8.5\pm1.42$& $13.56\pm0.122$\\
9& $3.390717$& $40$& $10.3\pm1.78$& $13.87\pm0.082$\\
10& $3.390960$& $57$& $8.3\pm1.68$& $13.55\pm0.156$\\
11& $3.391181$& $72$& $9.4\pm1.09$& $13.52\pm0.084$\\
12& $3.391497$& $93$& $16.2\pm0.87$& $13.48\pm0.031$\\
\hline
\multicolumn{5}{c}{\textbf{UVES}}\\
1& $3.388094$& $-139$& $11.2\pm0.11$& $13.95\pm0.005$\\
2& $3.388660$& $-100$& $43.3\pm1.05$& $13.84\pm0.015$\\
3& $3.388702$& $-98$& $4.3\pm0.42$& $12.86\pm0.028$\\
4& $3.389022$& $-76$& $14.3\pm0.68$& $13.31\pm0.029$\\
5& $3.389681$& $-31$& $4.0\pm0.22$& $13.06\pm0.015$\\
6& $3.390156$& $2$& $1.6\pm0.66$& $12.89\pm0.076$\\
7& $3.390182$& $4$& $22.7\pm0.64$& $14.09\pm0.013$\\
8& $3.390781$& $44$& $19.7\pm0.83$& $14.10\pm0.019$\\
9& $3.391200$& $73$& $11.5\pm0.68$& $13.54\pm0.043$\\
10& $3.391548$& $97$& $12.8\pm0.49$& $13.38\pm0.021$\\
\end{tabular}
\end{center}
\end{table*}

\subsection{Comparison of DLA properties between ESPRESSO and UVES}

Table \ref{tab:Ntot} provides the total column densities derived from the ESPRESSO and UVES Voigt profile fitting analysis. Errors on the total column densities include continuum fitting errors. Comparing the total column densities derived from ESPRESSO with those from UVES \citep[both from the analysis in this paper and ][]{Molaro00}, all values are consistent. We note that \cite{Molaro00} only fitted the low ions absorption profile as a single component profile, thus excluding the weak components 1, 2, and 5 from the ESPRESSO fit (Table \ref{tab:loComp}; components 1 and 4 from UVES). However, we emphasize that these three weak components contribute minimally to the total column densities, and our derived total column densities are consistent with \cite{Molaro00}. Thus, while in Table \ref{tab:Ntot} we conservatively assume logN(Si\ion{ii}) is a lower limit as components 1 and 2 are severely blended, it is likely that they have little contribution to the total column density of Si\ion{ii}. Ignoring these missing components would imply a total logN(Si\ion{ii})~$=14.99\pm0.217$ from ESPRESSO.

\begin{figure}
    \includegraphics[width=0.5\textwidth]{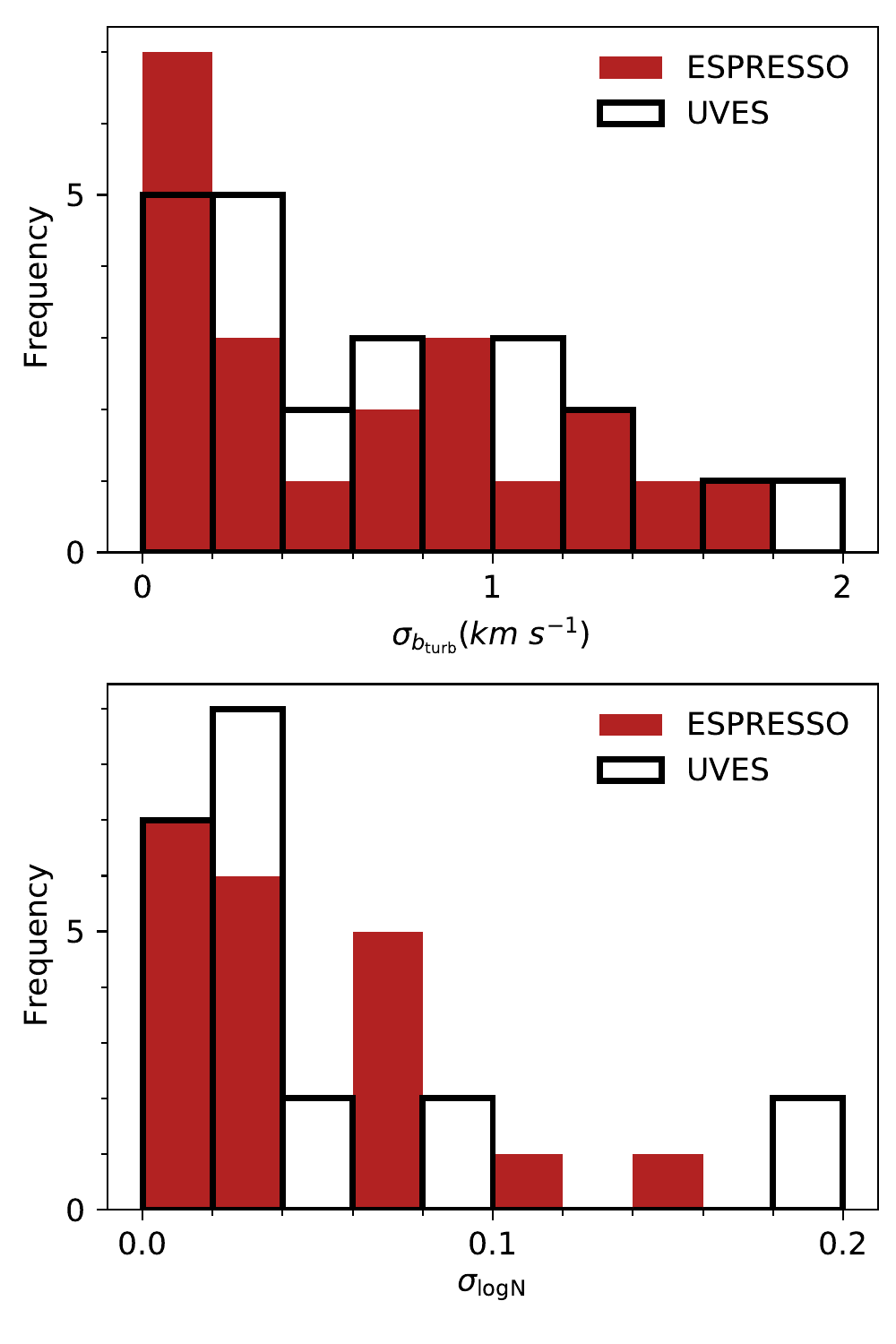}
    \caption{Distributions of the errors on the fitted components' \bturb{} (top panel) and logN (bottom panel) for ions Al\ion{ii}, Fe\ion{ii}, and C\ion{iv} from ESPRESSO (solid red bars) and UVES (hollow black bars) fits assuming the same number of components as the ESPRESSO fits for the respective ions (Tables \ref{tab:loComp} and \ref{tab:hiComp}). The UVES distributions show the errors derived from fitting the same redshift components as used in the ESPRESSO fits in order to provide an equal comparison. Despite the lower S/N in the ESPRESSO data, the errors on the component fits for ESPRESSO are similar in magnitude as those of UVES.} 
    \label{fig:errs}
\end{figure}

Despite the $\approx2\times$ lower S/N in the ESPRESSO data, the errors on the ESPRESSO total ion column densities are only $\approx2\times$ higher than those of UVES. However, we emphasize that the number of components used in the fitting of the low ions and C\ion{iv} profiles are not identical, preventing a fair comparison of the errors due to the degeneracy in component structure. Taking the redshifts of the fitted components for the ions Al\ion{ii}, Fe\ion{ii}, and C\ion{iv} from ESPRESSO and fixing them in the UVES profile fitting results in both \bturb{} and logN being generally consistent when the same number of components are used in the fitting (Tables \ref{tab:NloFixEsp} and \ref{tab:NhiFixEsp}; for low ions and C\ion{iv} respectively). We exclude Si\ion{ii} from the comparison due to the artefact seen in the UVES data. Figure \ref{fig:errs} shows the distribution of errors on \bturb{} (top panel) and logN (bottom panel) for this alternative fit. While the errors on the UVES fits typically vary  between $\approx 0.25-4.0\times$ that of ESPRESSO, the distribution of errors are comparable in size between the two datasets. Therefore, the combination of lower S/N but higher resolution does not significantly change the precision of fitting the same absorption line profile model. However, the combination of complex absorption profiles with either the lower S/N of the ESPRESSO data or lower resolution of UVES implies that the parameter space of the fitting process becomes highly degenerate and thus $\chi^2$ minimization techniques may not converge on similar profiles fitted.

\begin{table*}
\small
\begin{center}
\caption{Total column densities}
\label{tab:Ntot}
\begin{tabular}{lcccccc}
\hline
Instrument& logN(C\ion{iv})& logN(N\ion{i})& logN(Si\ion{ii})& logN(Al\ion{ii})& logN(Fe\ion{ii})& logN(Ni\ion{ii})\\
\hline
ESPRESSO& $14.70\pm0.021$& $14.63\pm0.015$& $>15.01$& $13.44\pm0.040$& $14.75\pm0.018$& $13.42\pm0.060$\\
UVES& $14.71\pm0.007$& \nodata{}& $>15.12$& $13.47\pm0.022$& $14.76\pm0.015$& \nodata{}\\

\end{tabular}
\end{center}
\end{table*}

The two sets of data span two decades and we checked for a temporal evolution in the column density of the individual profiles' components by refitting the UVES Fe\ion{ii} and Al\ion{ii} lines with the same five components used to fit the ESPRESSO data. While keeping $z$ and \bturb{} fixed, logN was refit for the UVES data (Tables \ref{tab:Nevolve}). No significant evolution in column densities was detected. Within $1\sigma$, all logN(Fe\ion{ii}) for each fitted component in the UVES data were consistent with the equivalent component in ESPRESSO (Table \ref{tab:loComp}), while only three components of the saturated Al\ion{ii} line were consistent within $1\sigma$ (the other two were within $2\sigma$). We therefore cannot confirm potential changes in the column densities over the past two decades.

\begin{table*}
\small
\begin{center}
\caption{Fitted UVES Voigt profile parameters for low ions using the same five components' redshifts of the ESPRESSO fit from Table \ref{tab:loComp}}
\label{tab:NloFixEsp}
\begin{tabular}{lccccc}
\hline
Comp.& $z$& $\Delta v$ (km s$^{-1}$)& \bturb{} (km s$^{-1}$)& logN(Al\ion{ii})& logN(Fe\ion{ii})\\
\hline
1& $3.389512$& $-42$& $8.5\pm0.27$&  $12.12\pm0.017$& $13.39\pm0.017$\\
2& $3.389892$& $-16$& $4.3\pm0.12$&  $12.49\pm0.018$& $13.60\pm0.019$\\
3& $3.390056$& $-5$& $6.0\pm0.16$&  $12.97\pm0.035$& $14.42\pm0.022$\\
4& $3.390188$& $4$& $4.6\pm0.11$&  $12.89\pm0.036$& $14.19\pm0.021$\\
5& $3.390481$& $24$& $15.2\pm1.13$&  $12.05\pm0.033$& $12.81\pm0.086$\\
\end{tabular}
\end{center}
\end{table*}

\begin{table}
\small
\begin{center}
\caption{Fitted UVES Voigt profile parameters for C\ion{iv} using the same twelve components' redshifts of the ESPRESSO fit from Table \ref{tab:hiComp}}
\label{tab:NhiFixEsp}
\begin{tabular}{lcccc}
\hline
Comp.& $z$& $\Delta v$ (km s$^{-1}$)& \bturb{} (km s$^{-1}$)& logN(C\ion{iv})\\
\hline
1& $3.388100$& $-139$& $11.2\pm0.11$& $13.95\pm0.004$\\
2& $3.388645$& $-101$& $43.8\pm1.05$& $13.84\pm0.014$\\
3& $3.388729$& $-96$& $4.3\pm0.43$& $12.86\pm0.028$\\
4& $3.389040$& $-74$& $14.1\pm0.68$& $13.29\pm0.029$\\
5& $3.389703$& $-29$& $5.7\pm0.22$& $13.25\pm0.012$\\
6& $3.389918$& $-15$& $5.4\pm0.64$& $13.17\pm0.083$\\
7& $3.390172$& $3$& $11.5\pm1.28$& $13.90\pm0.046$\\
8& $3.390453$& $22$& $8.2\pm1.61$& $13.41\pm0.184$\\
9& $3.390717$& $40$& $16.2\pm1.84$& $14.06\pm0.044$\\
10& $3.390960$& $57$& $4.7\pm1.22$& $12.98\pm0.186$\\
11& $3.391181$& $72$& $10.3\pm0.69$& $13.59\pm0.037$\\
12& $3.391497$& $93$& $13.8\pm0.52$& $13.43\pm0.020$\\
\end{tabular}
\end{center}
\end{table}

\begin{table*}
\small
\begin{center}
\caption{Fitted UVES component column densities using fixed ESPRESSO \bturb{} and $z$ from Table \ref{tab:loComp}}
\label{tab:Nevolve}
\begin{tabular}{lccccc}
\hline
Ion & logN$_{\rm Comp.~1}$ & logN$_{\rm Comp.~2}$ & logN$_{\rm Comp.~3}$ & logN$_{\rm Comp.~4}$ & logN$_{\rm Comp.~5}$ \\
\hline
Al\ion{ii} & $12.15\pm0.013$ &$12.31\pm0.018$ &$13.18\pm0.068$ &$12.90\pm0.026$ &$12.06\pm0.018$ \\
Fe\ion{ii} & $13.37\pm0.018$ &$13.60\pm0.015$ &$14.59\pm0.024$ &$14.08\pm0.017$ &$12.99\pm0.060$ \\
\end{tabular}
\end{center}
\end{table*}

\section{Conclusions}

This paper presents an analysis of the performance of ESPRESSO's high spectral resolution, $4\times2$ binning mode (HR42) for observing faint quasars. Using four hours of on-target integration, we observed the quasar QSO J0003$-$2603 (AB magnitude $R = 18.5$~mag). An archival spectrum of the same quasar obtained with UVES for approximately the same exposure time allows one to compare ESPRESSO's performance at extracting properties of an intervening DLA absorber. Despite having $\approx2\times$ smaller S/N (per pixel) than the UVES spectra (Figure \ref{fig:snr}), the ESPRESSO HR42 mode observations were able to recover column densities and absorption profiles to nearly the same precision as the UVES data. For complex absorption profiles with many overlapping narrow features (such as the typical C\ion{iv} profiles seen in strong Ly$\alpha$ absorbers), fitting Voigt profiles becomes degenerate with the modest S/N of our ESPRESSO data. The low ESPRESSO throughput below $\lambda \approx5000$~\AA{} prevents accurate continuum fitting and modelling of lines. The higher resolving power of ESPRESSO ($R\approx136\,700$) allowed to identify additional narrow (i.e., small turbulent component of the Doppler broadening parameter) substructure within the metal absorption profiles that were missed by the typical UVES setup used for observing DLAs ($R\approx48~000$) despite the $\approx2\times$ lower S/N (per pixel). Whilst UVES can achieve a resolving power beyond 100~000, in order to match the same S/N depth  as ESPRESSO (after accounting for differing pixel size), observations with UVES would require double the exposure time. Furthermore, the spectral resolution element of UVES would not be properly sampled at this resolution. Thus ESPRESSO's HR42 mode provides an optimal combination of extended wavelength coverage, resolving power, and observing efficiency above 5000~\AA{} that UVES cannot obtain.

In summary, ESPRESSO's HR42 mode can provide additional stability and $\approx3\times$ the spectral resolving power with a comparable performance to UVES beyond $\lambda \geq 5000$ \AA{} for analysing quasar absorption line systems. As the equivalent width of a line scales as FWHM divided by S/N, the HR42 mode of ESPRESSO would be an ideal avenue for searching for weaker lines, such as C\ion{i}, Si\ion{i}, and rare elements, typically not easily identified with UVES. The increased resolution and sensitivity of the HR42 mode combined with the stability and precise wavelength calibration of ESPRESSO enables precision physics and cosmological experiments \citep{Irsic17, Cooke20, Schmidt21}, and observations quantifying isotopic ratios and gas temperatures of individual absorbers \citep{Narayanan06, Cooke15, Welsh20, Noterdaeme21} whilst minimizing observing time requirements using a single UT.

\begin{acknowledgements}
We are grateful for assistance, discussions and comments provided by Ryan Cooke, Valentina D'Odorico, Alain Smette, and Louise Welsh that significantly improved this manuscript, and for the comments from the anonymous referee that helped improve the clarity of the text. We are thankful for the ESPRESSO observations that were supported by and collected at the European Southern Observatory under ESO programme 60.A-9801(W).
\end{acknowledgements}

\bibliographystyle{aa}
\bibliography{bibref}

\end{document}